\documentclass[aps,preprint,showpacs]{revtex4}
\usepackage{amssymb}
\usepackage{amsmath}
\usepackage{graphicx}
\usepackage{subfigure}
\usepackage{amsfonts}
\usepackage{color}
\begin{document}
\title{Effects of strain on the electronic structure of VO$_2$}

\author{Bence~Lazarovits$^{1,2}$, Kyoo~Kim$^1$, Kristjan~Haule$^1$, and Gabriel~Kotliar$^1$}
\affiliation{
$^1$Department of Physics, Rutgers University, Piscataway, New Jersey 08854, USA\\
$^2$Research Institute for Solid State Physics and Optics
of the Hungarian Academy of Sciences,
Konkoly-Thege M. \'{u}t 29-33., H-1121 Budapest, Hungary
}

\begin{abstract}
We present cluster-DMFT (CTQMC) calculations based on a downfolded tight-binding model 
in order to study the electronic structure of vanadium dioxide (VO$_2$) 
both in the low-temperature ($M_1$) and high-temperature (rutile) phases.
Motivated by the recent efforts directed towards tuning the physical properties
of VO$_2$ by depositing films on different supporting 
surfaces of different orientations
we performed calculations for different geometries for both phases. 
In order to investigate the effects of the different growing geometries we applied 
both contraction and expansion for the lattice parameter along 
the rutile $c$-axis
in the 3-dimensional translationally 
invariant systems miming the real situation.
Our main focus is to identify the mechanisms governing
the formation of the gap characterizing the $M_1$ phase
and its dependence on strain.
We found that the increase of the band-width with compression 
along the axis corresponding to the rutile $c$-axis
is more important than the Peierls bonding-antibonding 
splitting.

\end{abstract}

\maketitle

\section{Introduction}

Vanadium oxides compounds exhibiting exotic transport phenomena
are subjects of extensive interest. 
In particular vanadium dioxide, VO$_2$, undergoes a first-order transition from a
high-temperature metallic phase to a low-temperature insulating
phase at almost the room temperature ($T=340\,$K) \cite{morin}. 
There are intensive efforts around the world 
to make devices such as switches, transistors, detectors, varistors,
phase change memory, exploiting the unique properties of VO$_2$.

\begin{figure}[!bt]
\centering{
\includegraphics[height=0.5\linewidth]{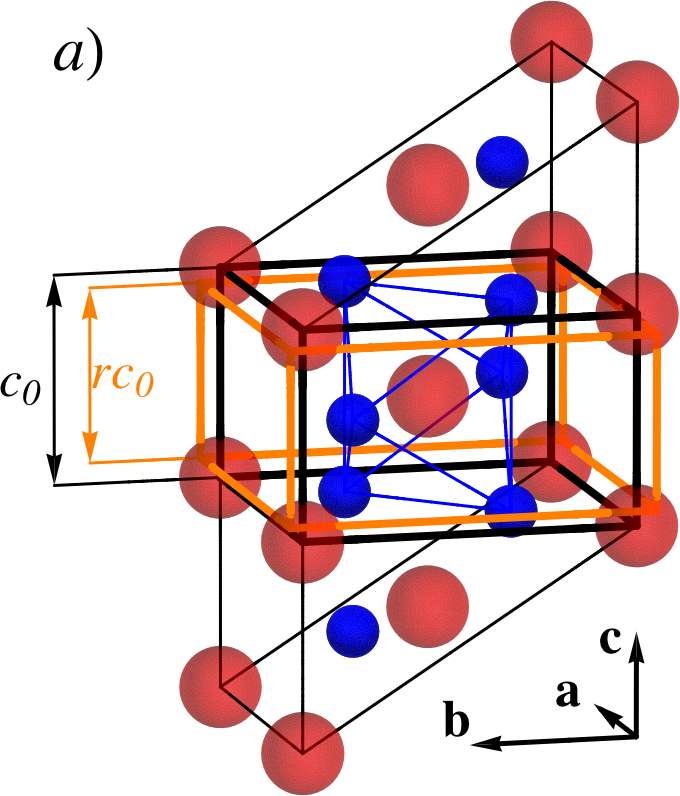}
\includegraphics[height=0.5\linewidth]{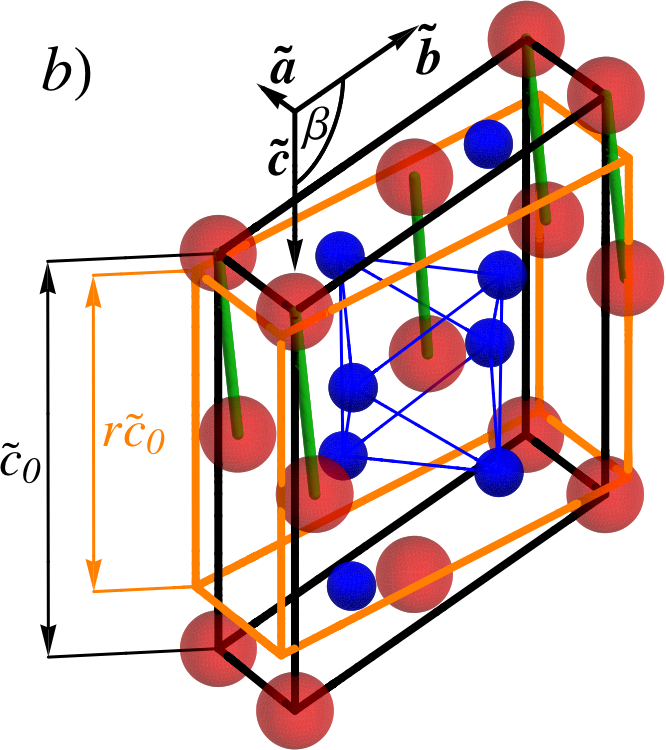}
  }
\caption{(Color online) a): The high temperature metallic rutile structure.
b): The low temperature insulating $M_1$ structure. 
{\em Large (red) spheres:} V atoms, {\em small (blue) spheres:} O atoms.
Thin black box in the rutile structure emphasizes 
the similarity between the two phases.
{\em Thick black boxes:} the unstressed unit cells, 
{\em thick gray (orange) boxes:} stressed unit cell. 
{\em Arrows:} the directions of lattice vectors (note the names
of the axes differ from the usual notation \cite{E02}).
The heights of the unstressed (stressed) unit cells are
$c_0$ ($c=r c_0$) for the rutile and $\tilde{c}_0$ ($\tilde{c}=r 
\tilde{c}_0$) for the $M_1$ structure. 
Green lines connect the dimerized V atoms.
The actual stress in
theory and experiments is 10 times smaller than depicted in
the figure. } 
\label{fig:structure}
\end{figure}

%%%%%%%%%%%%
% Geometry %
%%%%%%%%%%%%
At low-temperature VO$_2$ has a simple monoclinic ($M_1$) structure with space
group $P2_1/c$ ($M_1$~phase) while at high temperature it has a simple
tetragonal lattice with space group $P4_2/mnm$ rutile ({\em R}-phase) as
displayed in Fig.~\ref{fig:structure}.  The lattice structures of the
two phases are closely related as emphasized in
Fig.~\ref{fig:structure} by showing similar wedges of the different
lattices: the $M_1$ unit cell is similar to the {\em R} unit cell,
when the latter is doubled along the rutile $c$-axis.  For the sake of
simplicity, we will use the notation $c$-axis for both the $M_1$ and
{\em R} phase when referring to the axis equivalent to the rutile
$c$-axis (in $M_1$ phase this axis is sometimes called $a$-axis).  The
$M_1$ phase is characterized by a dimerization of the vanadium atoms
into pairs, as well as a tilting of these pairs with respect to the
$c$-axis \cite{KL70,andresson} as indicated by showing bonds between
the paired V atoms in Fig.~\ref{fig:structure}b.

% Arguably, it is due to this one-dimensional Peierls type distortion
% along the $c$-axis of the $M_1$ structure that
% the electronic and transport
% properties are dramatically different for the two phases.
%

The electronic and transport properties are dramatically different for
the two phases. The resistivity jumps by several orders of magnitude
through the phase transition, and the crystal structure changes from
{\em R}-phase at high-temperature to monoclinic $M_1$-phase at
low-temperature \cite{morin,allen}.  While the rutile phase is a
conductor, the $M_1$ phase is an insulator with a gap of $\sim0.6$~eV
\cite{KHH+06} at the Fermi energy.  VO$_2$ has also attracted a great
deal of attention for its ultrafast optical response, switching
between the {\em R} and the $M_1$ phases
\cite{lysenko,cavalleri,Baum}.  Despite the large number of
experimental
\cite{Baum,MH02,Maekawa,Qazilbash1,Qazilbash2,Qazilbash3,QBW+08,Eguchi,Ruzmetov,Braichovich,Dmitry,KHH+06}
and theoretical
\cite{BPL+05,Tomczak1,Tomczak2,Sakuma,Gatti,E02,RSA94,LCM06,LIB05}
studies focusing on this material the physics driving this phase
transition and the resulting optical properties are still not
identified undoubtedly.  In the theoretical works using LDA
\cite{E02,RSA94,allen} the formation of the gap was not found. The
single site DMFT approach~\cite{LCM06,LIB05} is known to correctly
describe the metallic phase but can not take into account the
formation of the bonding states of the dimers in the insulating phase
which requires the cluster DMFT method
\cite{BPL+05}.  As shown in Ref.~\cite{Tomczak2} the fully interacting
one-electron spectrum of the $M_1$ phase can be reproduced by an
effective band structure description.  The optical conductivities
calculated by using cluster DMFT shows good agreement with the
experiment for both phases~\cite{TB09}.

The metal-insulator transition (MIT) of VO$_2$ is usually attributed
to two different physical mechanisms.  One of
them is the Peierls physics, i.e. the dimerization of the V atoms
along the rutile $c$-axis \cite{g60}, and consequently opening of the
gap in the Brillouin-zone reduced in the direction along the
$c$-axis. In this case, the gap opening can be explained in the
framework of effective one-electron theories.
The other one is the Mott mechanism where the gap opens due to the
strong Coulomb repulsion between the localized V-$3d$ orbitals
\cite{ZM76} and the related dynamical effects.  Understanding in
detail the interplay and relative importance of both the Peierls and
the Mott mechanisms for the electronic structure is crucial for
controlling this material with an eye towards applications.  For
example, whether the driving force of this transition is electronic
(i.e. occurring on femtosecond timescales) or structural (occurring on
the picosecond timescale) is important to understand the speed of the
switching from the $M_1$ to the rutile phase.  From the perspective of
applications, in order to control the properties of vanadium dioxide,
it is essential to identify the effects of compressive and tensile stress
resulting from the various substrates, on which the films are
deposited \cite{MH02,Maekawa}.

The experimental results of Ref.~\cite{MH02} showed that compressive
uniaxial strain (along the $c$-axis) stabilizes the metallic phase.
This result cannot be explained by applying a simple Peierls picture
exclusively. The Peierls mechanism predicts that a compression along
the $c$-axis increases the splitting between the bonding and
antibonding state, formed by the combinations of particular $3d$
orbitals residing on the different V atoms in a dimer, and hence would
promote insulating state.
The Peierls mechanism thus increases the tendency to open 
the gap in the $M_1$
phase, which would stabilize the insulating phase at the expense of
the metallic phase. In this picture, the transition temperature
would increase under uniaxial compressive strain, opposite to
experiment~\cite{MH02}.

The LDA calculations fail to reproduce the gap in the $M_1$
phase~\cite{E02}.  This suggests that the correlations are important
in this material.
% 
% formation of the gap is a
% consequence of the interplay between Peierls and dynamical
% electron-correlation effects \cite{BPL+05}.
% 
Motivated by this and the above mentioned experiment we have examined
the influence of strain on the electronic structure of VO$_2$
microscopically applying the dynamical mean-field theory \cite{GKK+96}
combined with the local density approximation of density functional
theory (LDA+DMFT) \cite{KSH+06}.  We carried out LDA+DMFT calculations
of VO$_2$ under strain.  We established that in addition to the
Peierls distortion, other factors like the position of the $e_g^\pi$
band and the band-width of the bonding state in the $M_1$ phase
showing definite sensitivity to the strain, play a defining role in
driving the MIT in VO$_2$.  Theoretical predictions for the strain
dependence of many spectroscopic quantities are made, including the
photoemission, the optical conductivity, the inverse photoemission and
XAS in the $R$ and $M_1$ phases.  The insights achieved in this study
together with the computational machinery developed, will serve as a
basis for rational material design of VO$_2$ based applications.

\section{Method}

The unit cell of the rutile structure contains two vanadium and four
oxygen atoms while the $M_1$ structure contains four vanadium (2
dimers) and eight oxygen atoms.
The calculations in the rutile phase were performed with the doubled
unit cells (four V and eight O atoms) to allow the formation of a
bonding state between the V atoms separated along the $c$-axis, a
mechanism which has dramatic effect in the $M_1$ phase.
%mechanism which has dramatic affect in the $M_1$ phase.
%Hence almost identical many-body impurity problem was solved in both
%phases.
%
The lattice structure parameters for the rutile (referred as $a_0$,
$b_0$ and $c_0$ in the following) and $M_1$ cases ($\tilde{a}_0$,
$\tilde{b}_0$, $\tilde{c}_0$ and $\beta_0$) were published by McWhan
{\it et al.} in Ref.~\cite{MMR+74} and Kierkegaard and Longo in
Ref.~\cite{KL70}, respectively.  Note that our notation of the axis
differs from the published data \cite{E02}, in order to emphasize the
similarity of the {\em R} and $M_1$ structures.

The strain was applied by changing the lattice constants but not the
internal parameters of the atomic positions.  The strain is
characterized by the ratio between the experimental lattice constant
along the $c$-axis of the unstrained crystal, and the one used in the
calculation, $r=c/c_0$ in the rutile and $r=\tilde{c}/\tilde{c}_0$ in
the $M_1$ case.
The applied strains correspond to $r=0.98$, 
1.00 and 1.02 ratios both for the {\em R} and $M_1$ phases 
in order to make calculations which are compatible with 
the experiment in Ref.~\cite{MH02}. 
The lattice parameters were changed with the constraint 
to preserve the volume of the original unit cell and to form
a structure to closely resemble the original structure.
This is illustrated in Fig.~\ref{fig:structure} for both phases.
The constant volume constraint in the rutile phase requires for the
other two lattice constants $a=a_0/\sqrt{r}$ and $b=b_0/\sqrt{r}$.
For the $M_1$ structure, two other constraints were introduced:
the two ratios $(\tilde{b}\sin\beta)/\tilde{a}=0.9988$ and 
$-2(\tilde{b}\sin\beta)/\tilde{c}=1.0096$ were kept constant. Hence we
used the following lattice constants and angles
$\tilde{c}=r\tilde{c}_0$, 
$\tilde{a}= \tilde{a}_0 /\sqrt{r}$,
$\tan(\beta)= \tan(\beta_0) /r^{3/2} $ and 
$\tilde{b}=\tilde{b}_0\sqrt{r^2\cos^2(\beta_0)+\sin^2(\beta_0)/r}$.
%%%%%%%%%%%%%%%%%%%%%
% Electronic states %
%%%%%%%%%%%%%%%%%%%%%

\begin{figure}[!bt]
\begin{center}$
\begin{array}{ccc}
\includegraphics[height=0.3\linewidth,clip=true, viewport=0pt 40pt 500pt 500pt]{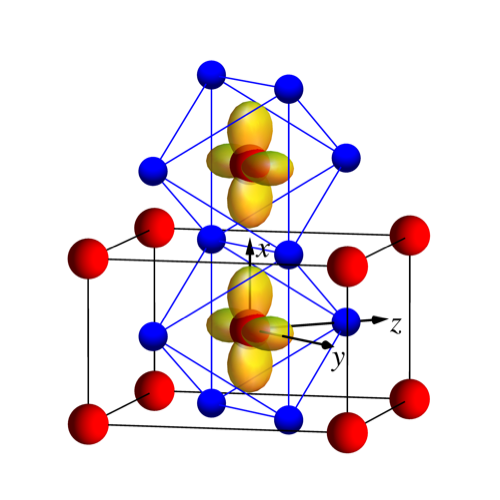}&
\includegraphics[height=0.3\linewidth,clip=true, viewport=0pt 40pt 500pt 500pt]{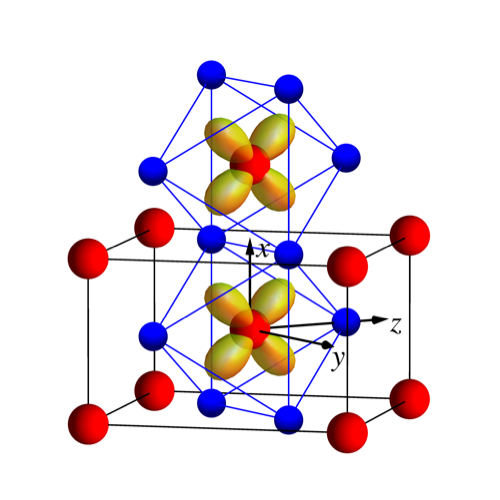}&
\includegraphics[height=0.3\linewidth,clip=true, viewport=0pt 40pt 500pt 500pt]{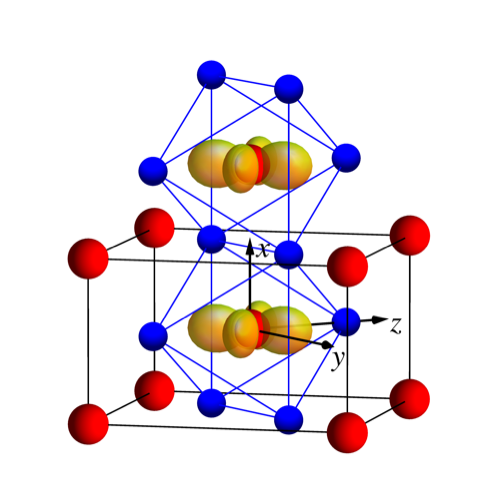}\\
d_{x^2-y^2}&d_{xz}&d_{yz}
\end{array}$
\end{center}
\caption{ 
(Color online)
The sketch of vanadium $t_{2g}$ orbitals 
($d_{x^2-y^2}$, $d_{xz}$, $d_{yz}$) together with the 
applied coordinate system and the V-V pairs considered in the rutile phase 
calculations.
{\em Large (red) spheres:} V atoms, {\em small (blue) spheres:}  O atoms. 
 } 
\label{fig:orbitals}
\end{figure}

For the {\em R}-phase, we used a local coordinate system introduced by
Eyert in Ref~\onlinecite{E02}, as shown in Fig.~\ref{fig:orbitals}.  In the
rutile geometry the $x$-axis is parallel to the rutile $c$-axis while
the $z$-axis points along the $[110]$ direction, pointing to the
adjacent oxygen atom in the vanadium plane.
In the $M_1$ phase, our local coordinate system is somewhat tilted to
align to monoclinic geometry.  The symmetry classification of the
electronic orbitals is also adopted, namely $d_{3z^2-r^2}$ and
$d_{xy}$ stand for the $e_g$ states and $d_{xz}$, $d_{yz}$ and
$d_{x^2-y^2}$ for the $t_{2g}$ orbitals.  The $t_{2g}$ manifold is
divided further: the notation $a_{1g}$ is used for $d_{x^2-y^2}$ state
and $e^{\pi}_g$ for the $d_{xz}$ and $d_{yz}$ states.  In this choice
of the local coordinate system the dimerization strongly affects the
$d_{x^2-y^2}$ states, which have large electron density along the V-V
bond.

\begin{figure}[!bt]
\centering{
\includegraphics[width=0.49\linewidth]{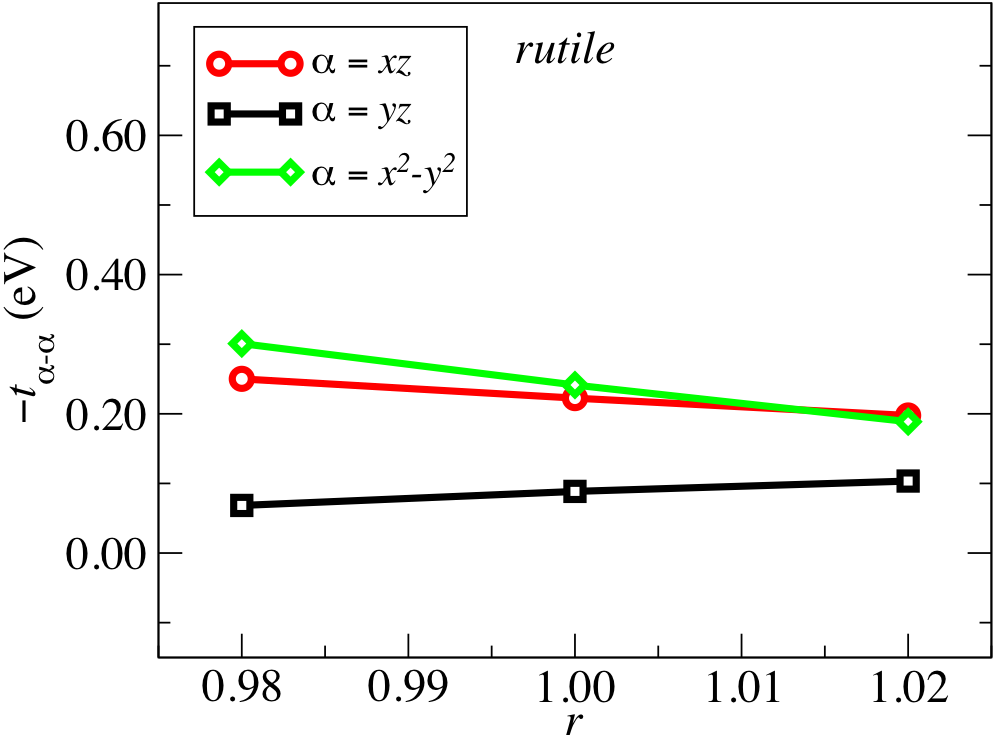}
\includegraphics[width=0.49\linewidth]{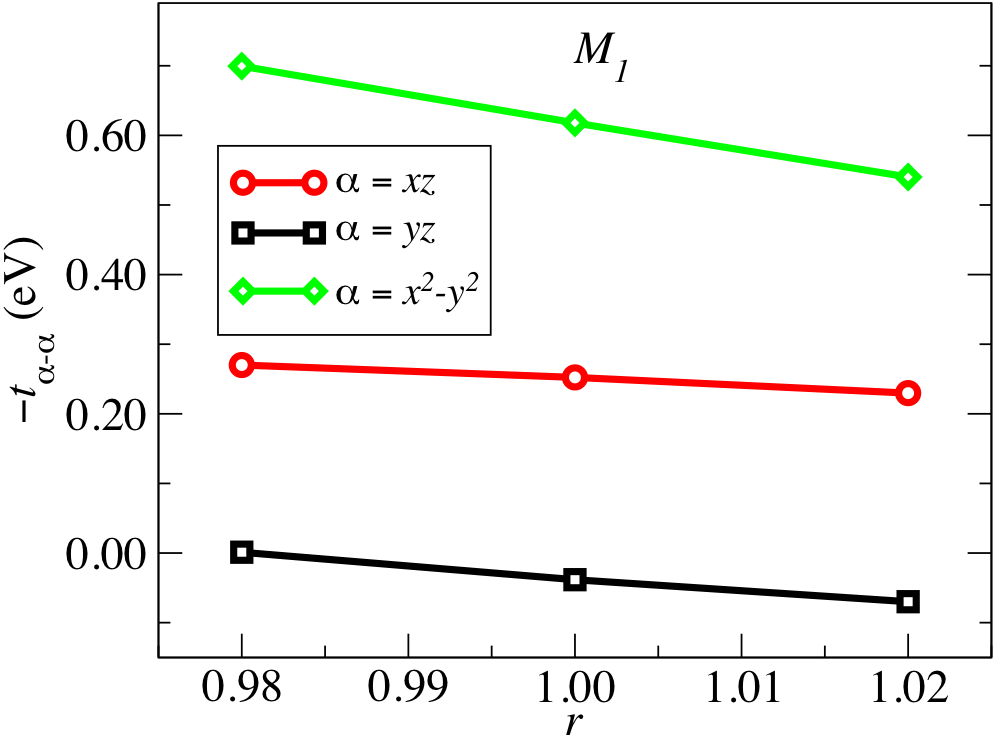}
  }
\caption{ 
(Color online)
  The intra dimer V-V hoppings 
  corresponding to the $t_{2g}$ orbitals shown in Fig.~\ref{fig:orbitals} 
  as a function of strain. The left (right) panel shows the parameters in the
  rutile ($M_1$) phase. The direction of strain is depicted in
  Fig.~\ref{fig:structure}.
}
\label{fig:hoppings}
\end{figure}

First, a self-consistent LDA calculation was performed using the
linear muffin-tin orbitals method combined with the atomic sphere
approximation (LMTO-ASA) \cite{lmto}.  For a satisfactory description
of the interstitial region 48 empty spheres were added in the LDA-ASA
calculations.  In the next step we determined a {\em downfolded
  parameters} of the low energy, effective Hamiltonian, $H_{eff}$,
including only the $3d$ $t_{2g}$ subset of electronic states
lying in the vicinity of the Fermi energy. The effective Hamiltonian
is constructed as:
\begin{equation}
H_{\mathrm{eff}}=
\sum\limits_{\sigma} 
\sum\limits_{\langle i,i'\rangle} 
\sum\limits_{ \alpha,\alpha'} 
\left(
\epsilon_{i,\alpha,\sigma} 
\delta_{\alpha,\alpha'}
\delta_{i,i'}
+
t_{i,\alpha,\sigma;i',\alpha',\sigma}
\right)
c_{i,\alpha,\sigma}^{\dagger}
c_{i',\alpha',\sigma}
\end{equation}
with off-diagonal hopping ($t$) and diagonal one-electron energy
($\epsilon$) parameters 
belonging to different sites $(i)$ and states 
($\alpha =d_{xz}$, $d_{yz}$, $d_{x^2-y^2}$) with a spin character $\sigma$. 
The applicability and accuracy of the above downfolding method 
is determined by the mutual 
positions and characters of the electronic bands. 
In the studied case of VO$_2$ both in the $M_1$ and {\em R} phases 
the bands in the proximity 
of the Fermi level have mainly $t_{2g}$ character and are well 
separated from both the 
low-lying $2p$-bands of the oxygen, and from the 
$e_g$ bands of vanadium, due to the strong crystal-field splitting 
($\sim 3.0-3.5$~eV \cite{E02}). The crystal field splitting is due
to oxygen octahedron  surrounding the individual V atoms. 
This mapping of the LDA band structure onto a tight-binding 
type Hamiltonian provides a reasonably good description of the electronic 
structure as compared to the LDA serving as a good starting point for the 
following DMFT calculations. 
The evolution of the intra dimer V-V hopping matrix elements 
corresponding to the three different $t_{2g}$ orbitals 
induced by the strain are shown in Fig.~\ref{fig:hoppings}.
Although, the relative changes in the parameters are small, the
manifestations of these changes in the low energy spectral function,
and therefore in the transport properties, are significant.

\begin{figure}[!bt]
\centering{
 \includegraphics[width=0.5\linewidth]{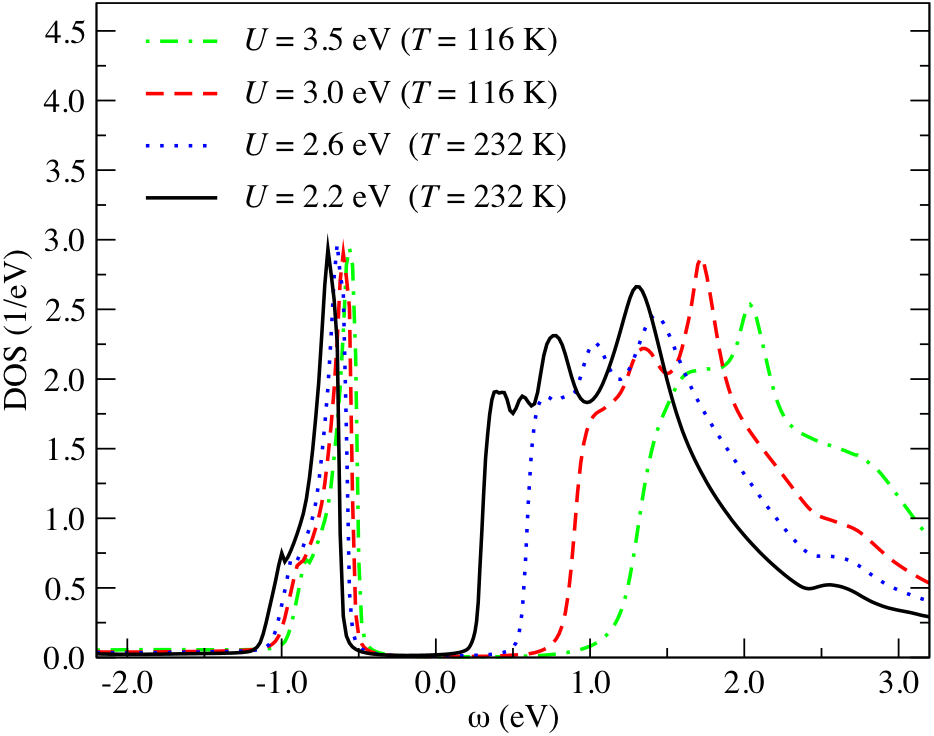}
  }
\caption{
(Color online)
The variation of the $t_{2g}$ density of states of the $M_1$ phase 
as a function of the applied Coulomb repulsion parameter, $U$. 
Note the high sensitivity of the gap on $U$.
}
\label{fig:var_U}
\end{figure}

%%%%%%%%
% DMFT %
%%%%%%%%
The effects of the electron-electron interaction beyond LDA are
treated by means of the dynamical mean-field theory
\cite{GKK+96,KSH+06}.  The coulomb interaction was taken into account
for the V-V dimers by using the cluster extension of the DMFT
\cite{KSH+06}.  In this approach the quantities describing the
electronic states of a cluster (Green's function, self-energy, {\it
  etc.}) are matrices of both the site and electronic state indices.
Having two sites (two V atoms) and three states ($t_{2g}$ states) the
resulted matrices have dimensions of $6\times6$ for non-magnetic
calculations.  For solving the impurity problem, we used the
continuous time quantum Monte Carlo method (CTQMC)~\cite{H07} at the
temperatures $T=232$~K and $T=390$~K, below and above the critical
temperature, for the $M_1$ and {\em R} cases, respectively.  In the
present calculations we assumed on-site Coulomb interaction on the V
atoms written as:
\begin{equation}
H_U=
U
\sum\limits_{i=1,2} 
\sum\limits_{ \alpha \alpha'} 
\sum\limits_{ \sigma \sigma'} 
c_{i,\alpha\sigma}^{\dagger}
c_{i,\alpha\sigma}
c_{i,\alpha'\sigma'}^{\dagger}
c_{i,\alpha'\sigma'}
(1-\delta_{\alpha\alpha'}
\delta_{\sigma\sigma'})
\end{equation}
excluding the terms with similar orbital and spin character at same
time.  We fixed the parameter $U$ at $U=2.2$~eV, to reproduce the
experimentally measured gap in the $M_1$ phase, which shows a weak
temperature dependence between 100 and 340~K, varying between
$\sim0.75$ and $\sim0.6$~eV \cite{OFO01,KHH+06}.  The strong
dependence of the gap size on the $U$ parameter is shown in
Fig.~\ref{fig:var_U}.
% In our calculations $J=0$ was taken disregarding
% the Hund coupling completely.

\begin{figure}[!bt]
\centering{
\includegraphics[width=0.5\linewidth]{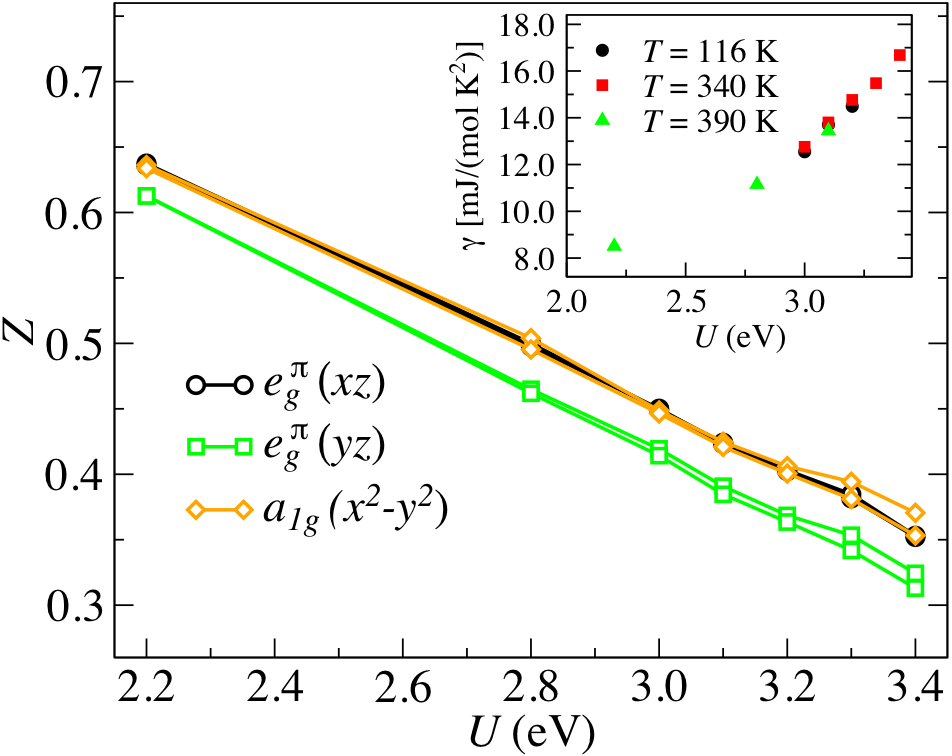}
  }
\caption{
(Color online)
Renormalization constants 
$Z_{\alpha}=(1-\partial \mathrm{Re}\Sigma_{\alpha}(\omega)/\partial \omega)^{-1}$
of the different orbitals in the {\em R} phase as a function of $U$. 
{\em Inset:} electronic specific heat, 
$\gamma=\sum\limits_{\alpha}\rho_{\alpha}(0)/Z_{\alpha}$, 
calculated at different temperature and values of $U$.
}
\label{fig:Z_gamma}
\end{figure}

The linear $U$ dependence of the renormalization factor $Z$ 
calculated from the real part of the self-energy of the 
different orbitals $(\alpha)$ as 
$Z_{\alpha}=(1-\partial \mathrm{Re}\Sigma_{\alpha}
(\omega)/\partial \omega)^{-1}$ 
and the electronic specific heat, 
$\gamma=\sum\limits_{\alpha}\rho_{\alpha}(0)/Z_{\alpha}$,
of the rutile phase are shown in Fig.~\ref{fig:Z_gamma}. 
The $Z$ value obtained at $U=2.2$~eV is in good agreement with the one
published in Ref.~\cite{BPL+05}.  An alternative method to determine
the value of $Z$ factor is to take the ratio of the experimentally
measured {\em plasma frequency} in the rutile phase
($\omega_p^{exp}=2.75$~eV)\cite{QBW+08} and devide it by the band
theoretical LDA-LAPW calculation \cite{wien2k}
($\omega_p^{LDA}\approx4.1$~eV), i.e.,
$\omega_p^{exp}/\omega_p^{LDA}\approx0.67$, which agrees well with
$Z\approx 0.62$ in the case of $U=2.2$~eV.  Although the $U$ was
chosen to reproduce the gap in the $M_1$-phase, it is satisfying that
this value is also compatible with the alternative estimation, based
exclusively on the optical and thermodynamic properties of the rutile
phase.
%
%It is necessary to mention that this value is smaller than the one
%considered in the paper of Biermann {\it et al} \cite{BPL+05} using a
%similar cluster DMFT approach with parameters $U=4.0$~eV and
%$J=0.68$~eV. It is mentioned in that work that with a small $J$ the
%insulator phase is stabilized at smaller $U$ which is in accordance
%with our results.  Besides the different treatment of the Hund
%coupling the differences between our results and the ones in
%Ref.~\cite{BPL+05} may can be attributed to the different impurity
%solver (Hirsch-Fye quantum Monte Carlo) and the different temperature
%(770~K) were applied in Ref.~\cite{BPL+05}.

In order to achieve a structured (almost diagonal) self-energy matrix
for the cluster, we used a basis of symmetric $(s)$ and anti-symmetric
$(as)$ combination of the states localized on the individual V atoms
of the dimers defined as
\begin{equation}
(c_{\alpha,\sigma }^{s(as)})^\dagger
=
\frac{1}{\sqrt{2}}(c_{1,\alpha,\sigma}^{\dagger} \pm c_{2,\alpha,\sigma}^{\dagger})
,\quad
(\alpha \in t_{2g})
\quad.
\label{eq:b-ab}
\end{equation}
To obtain the physical properties at real energies we performed
analytic continuation to the real axis using a recently developed
method of expansion in terms of modified Gaussians and a
polynomial fit at low frequencies, as described in details in
Ref.~\cite{HYK09}.
%%%%%%%%%%%%%%%%%%%%%%%%%
% Computational details %
%%%%%%%%%%%%%%%%%%%%%%%%%

\section{Results and Discussion}

\begin{figure}[th!]
 \includegraphics[width=0.49\linewidth]{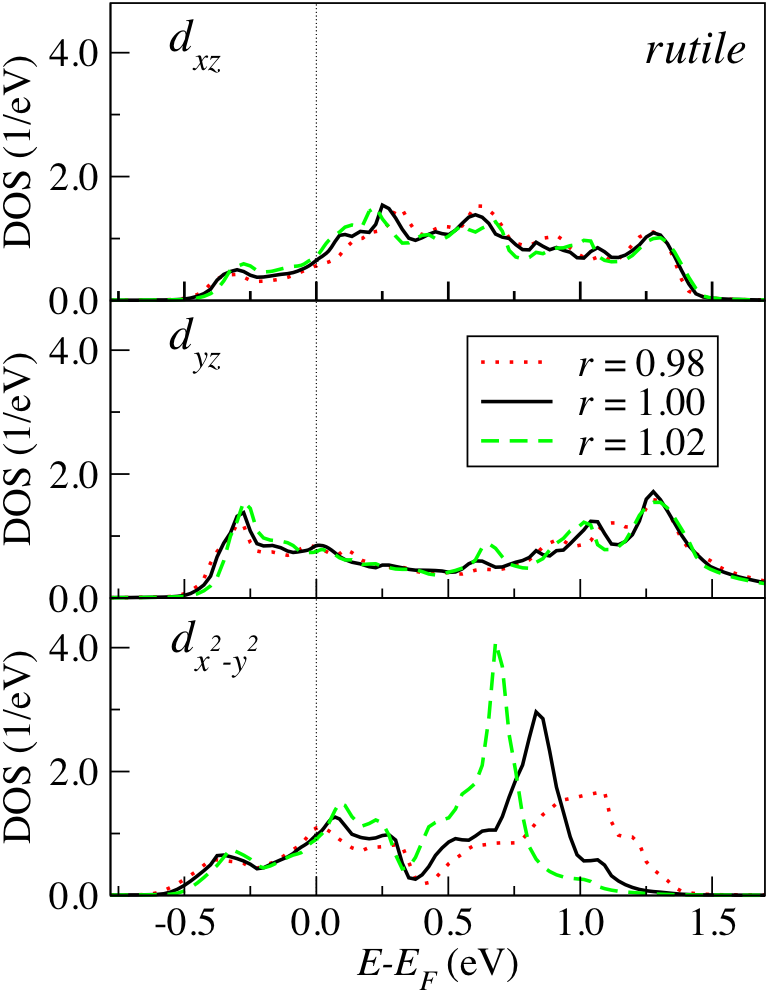}
 \includegraphics[width=0.49\linewidth]{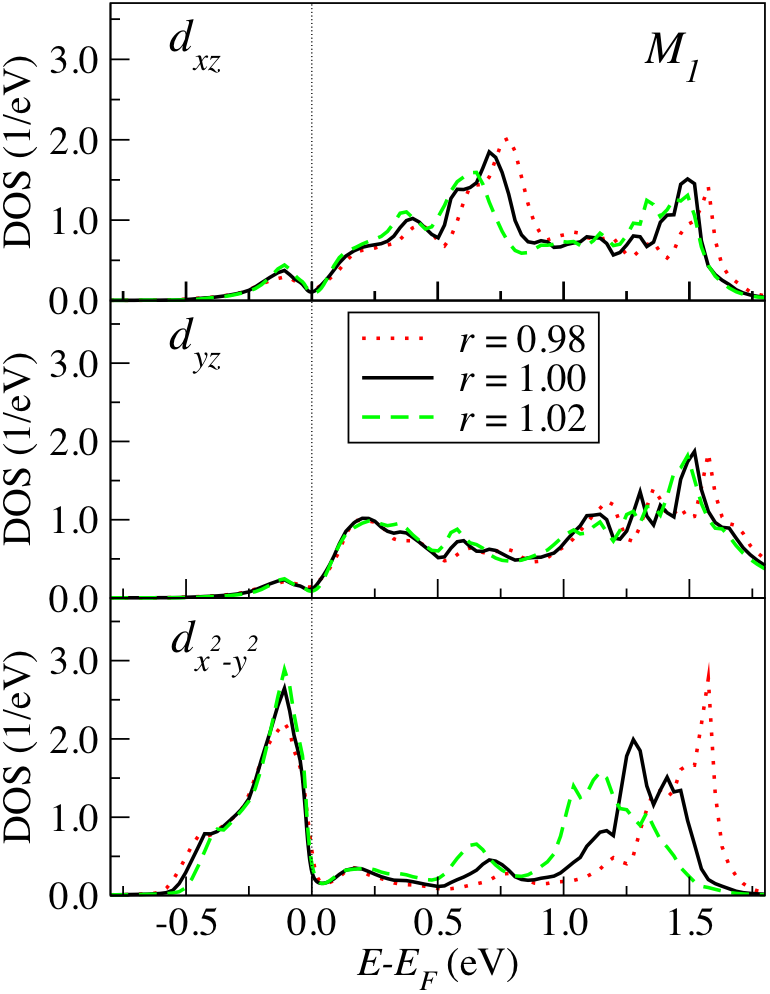}
\caption{
(Color online)
LDA partial DOS of the $t_{2g}$ states
of a V atom in the 
rutile (left) and $M_1$ (right) phases in the proximity of the Fermi 
energy 
with different $r$ ratios:
$r=0.98$ (dotted, red), 1.00 (full, black) and 1.02 (dashed, green). 
Note the large splitting of the $d_{x^2-y^2}$ states and the barley 
changed DOS at the Fermi level.
}
\label{fig:lda_dos}
\end{figure}

Fig.~\ref{fig:lda_dos} displays the LDA $t_{2g}$ densities of states
(DOS) of the V atoms in both the rutile and $M_1$ phases
for different $r$ ratios ($r=0.98$,~1.00,~1.02).
The total band-width ($\sim2.6$~eV for both phases) and also the fine
details of the DOS obtained by the LDA-LMTO
%and by the downfolded
%model also (not shown here)
agree well with previous studies \cite{E02,LCM+06}.  The minor
discrepancies are probably a consequence of the different electronic
structure method or the slightly different geometry.

The effect of the stress applied along the rutile {\it c}-axis on the
electronic DOS can be clearly seen.  The position of the $a_{1g}$ peak
is changed considerably, while the total band-width is only slightly
changed ($<0.1$~eV) under application of stress.  Only the
$d_{x^2-y^2}$ orbital changes significantly, while the 
two $e_{g}^\pi$ orbitals remain mostly unchanged.
This is due to the sensitivity of the overlap integral of
$d_{x^2-y^2}$ states along the $c$-axis.  It is clear from
Fig.~\ref{fig:lda_dos} that the splitting of the $d_{x^2-y^2}$ peaks
strongly increases with decreasing lattice parameter $c$ roughly
following the linear dependence of the intra-dimer hopping parameter
$t_{x^2-y^2,x^2-y^2}$ on ratio $r$, shown in Fig.~\ref{fig:hoppings}.
However, even under compressive stress of $r=0.98$, the splitting is
not large enough to open a gap in the $M_1$ phase.  The value of the
intra-dimer hopping parameters corresponding to the $d_{x^2-y^2}$
states, which plays a significant role in the formation of the
electronic structure of VO$_2$, are $-0.30$~eV and $-0.61$~eV for the
{\em R} and $M_1$ cases, respectively. These values are in a good
agreement with published values of Ref.~\onlinecite{BPL+05}.
One can observe that the splitting of the $d_{x^2-y^2}$ peaks are
larger for the $M_1$ geometry ($\sim1.4$~eV) than for the rutile
structure ($\sim0.76$~eV). This can be attributed to the reduced V-V
distance along the $c$-axis due to the dimerization.  The splitting of
the bands can be roughly approximated by $2t_{x^2-y^2,x^2-y^2}$ for
the different $r$ ratios.  This behavior resembles the bonding and
anti-bonding splitting of a dimer molecule, suggesting that the
splitting of these states is determined mainly by the intra dimer
hopping, especially in the $M_1$ case.  For the $e_g^\pi$ states, this
correspondence is less clear, showing the importance of the
inter-dimer hoppings.  It is worth to note that the LDA calculations
can capture only the Peierls physics, which alone is not sufficient to
explain the formation of the gap even if a reduced V-V distance is
considered.

To trace the effect of the electron-electron interactions {\it e.g.},
appearance of the gap in the $M_1$ phase and the reduction of the
band-width, we carried out LDA+DMFT calculations.  Our theoretically
calculated {\it orbitally resolved } spectral functions can be
directly compared to the measured angle integrated photoemission (PES)
and x-ray absorption spectroscopy spectra (XAS).  In the last few years a
large number of experimental studies of these kind were carried out
probing also the many body character of the occupied and the
unoccupied states in VO$_2$ and serving as a stringent test of the
theoretical approach.  This validation of our LDA+DMFT results is
crucial before proceeding to make reliable predictions for the
strained materials for which these spectroscopic information is not
yet available.

%%%%%%%%%%%%%%%%%%%%%%%%%%%%
% DMFT total upfolded DOS  %
%%%%%%%%%%%%%%%%%%%%%%%%%%%%
\begin{figure}[!bt]
\centering{
 \includegraphics[width=0.49\linewidth]{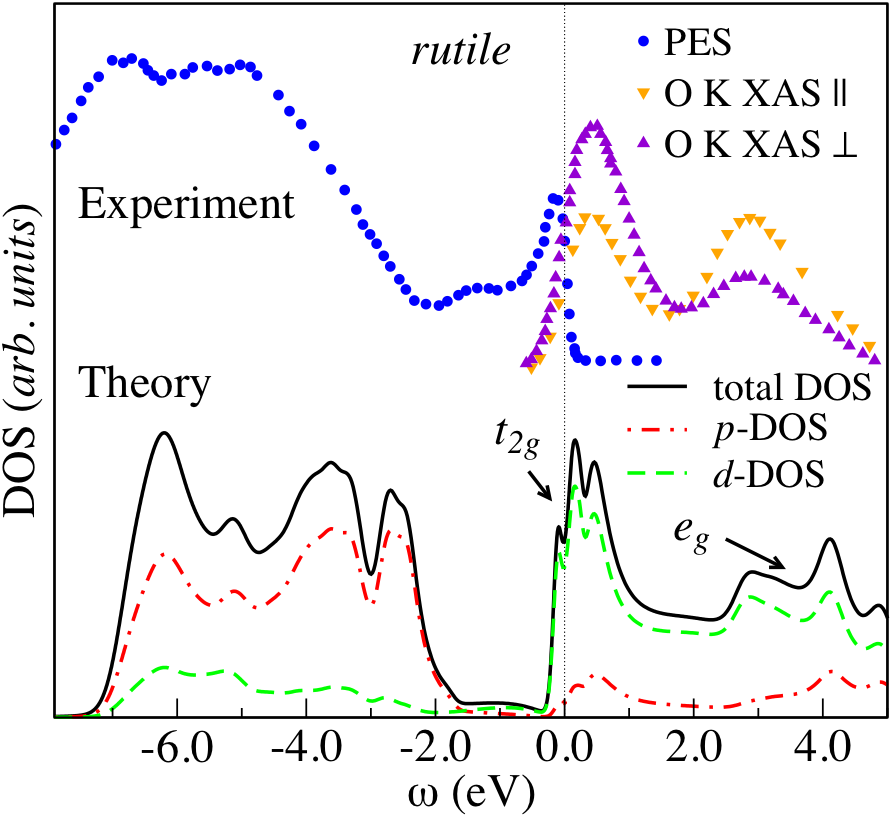}
 \includegraphics[width=0.49\linewidth]{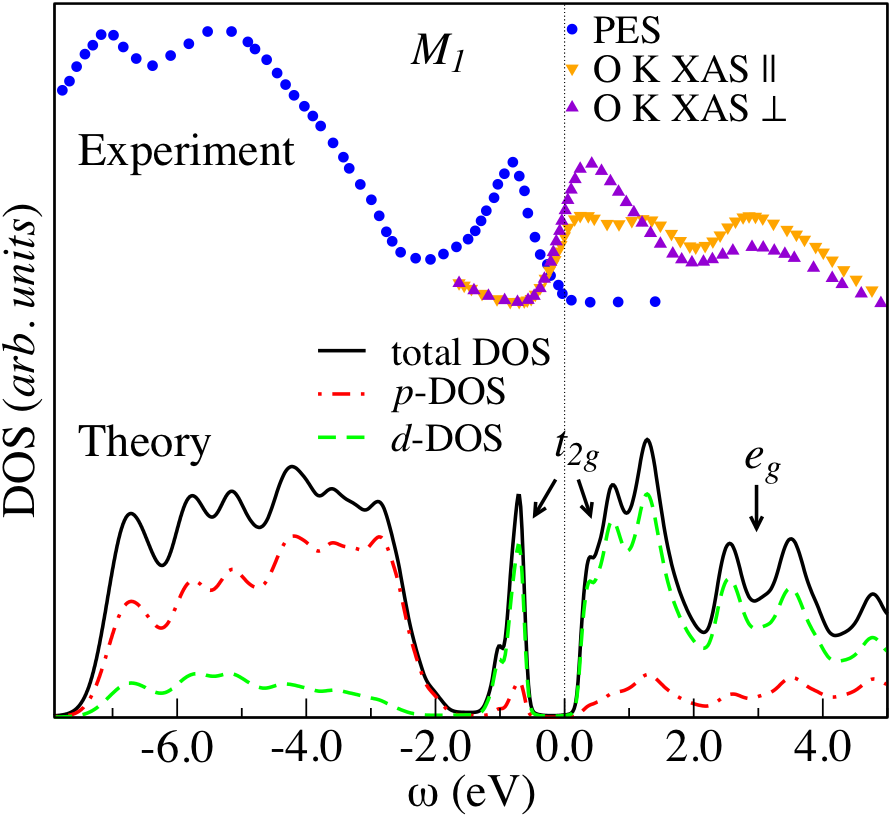}
}
\caption{ 
(Color online)
Comparison of V-$3d$, O-$2p$ and the total DOS of LDA+DMFT calculation 
in rutile (left) and $M_1$ (right) phases ($r=1.00$)
with angle integrated photoemission (PES)
and x-ray absorption spectroscopy spectra (XAS) measurement. 
The experimental results are reproduced from Ref.~\cite{KHH+06} and the 
XAS results were shifted to obtain the best agreement with our 
theoretical results.}
\label{fig:total}
\end{figure}

\begin{figure}[th!]
\includegraphics[width=0.4\linewidth]{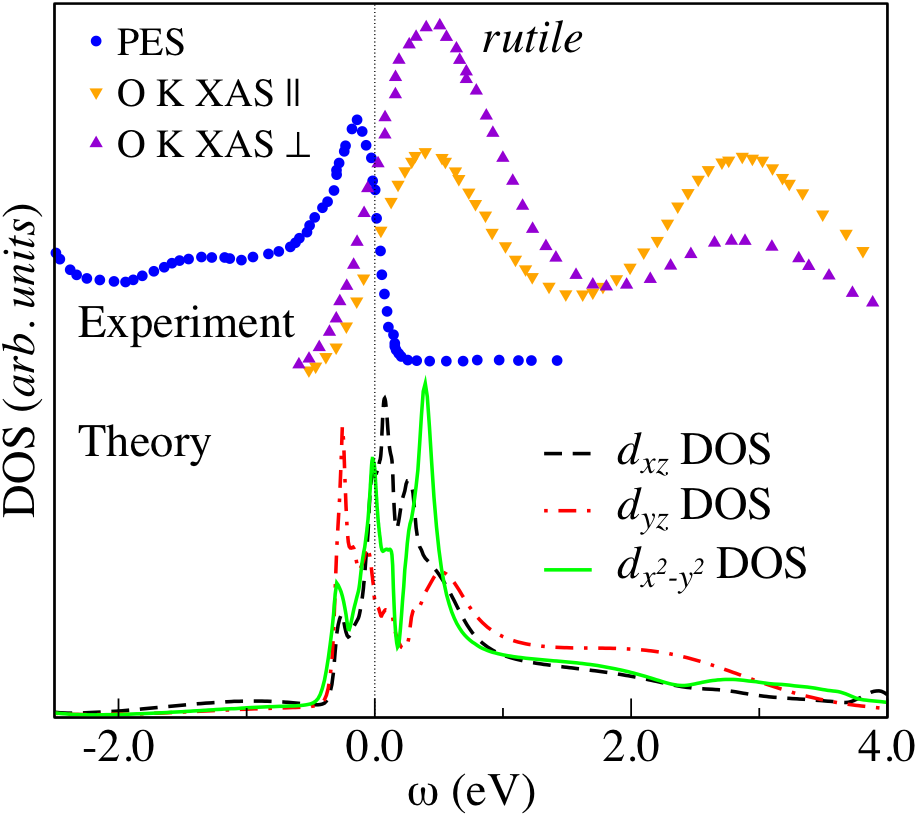}
\includegraphics[width=0.4\linewidth]{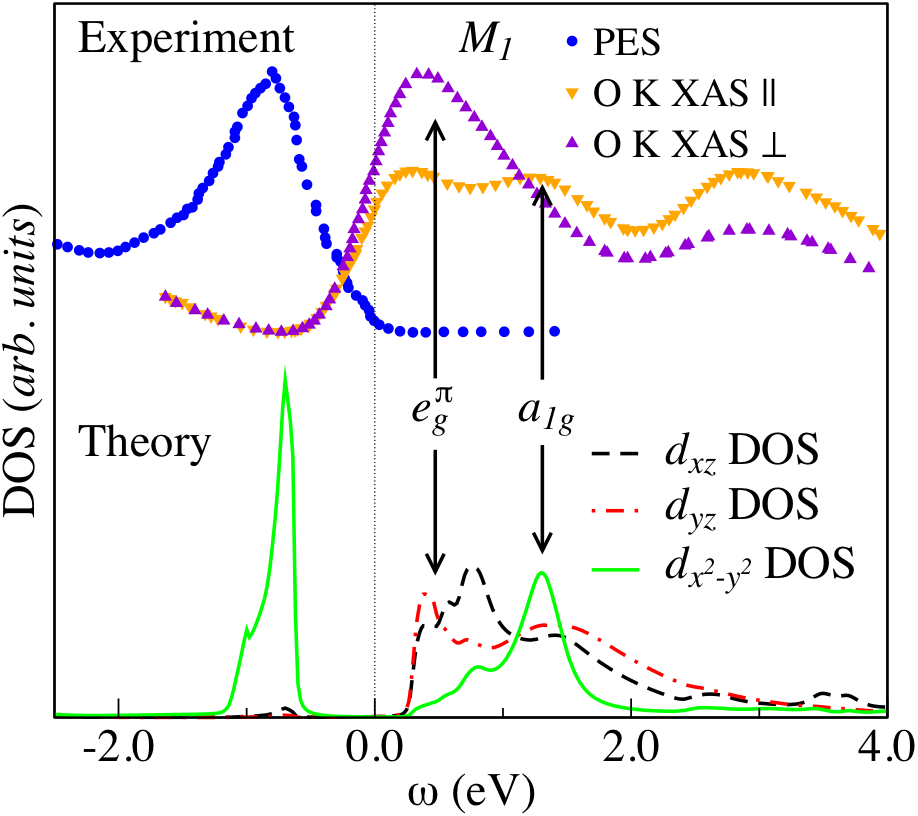}
\caption{
(Color online)
Comparison of the orbitally resolved $t_{2g}$ DOS of LDA+DMFT calculation 
corresponding to a V atom
in rutile (left) and $M_1$ (right) phases ($r=1.00$)
with angle integrated photoemission (PES)
and x-ray absorption spectroscopy spectra (XAS) measurement. 
The experimental results are reproduced from Ref.~\cite{KHH+06}.
The XAS result is shifted to obtain the best agreement with our
theoretical results.
}
\label{fig:resolved}
\end{figure}

The upfolded density of states of the LDA+DMFT, which includes besides
the V-$3d-t_{2g}$ states the O-$2p$ and the V-$3d-e_g$ states, are
shown in Fig.~\ref{fig:total}.
%%%%%%%%%%%%%%%%%%%%%%%%
% DMFT rutile full DOS %
%%%%%%%%%%%%%%%%%%%%%%%%
The dynamical correlation effects, not included in LDA, decrease the
width of the quasiparticle $t_{2g}$ states for $\sim 0.6$ times, in
agreement with our calculated quasiparticle renormalization amplitude
$Z\approx 0.62$.  In the rutile structure two Hubbard bands appear: a
very weak lower Hubbard band around -1.0~eV (was previously reported
in theoretical \cite{BPL+05} and experimental \cite{KHH+06} studies)
and a stronger upper hubbard band at 2.0~eV which overlaps with the
$e_g$ band (can be seen better in Fig.~\ref{fig:resolved} where the
$e_g$ band is not shown).
%%%%%%%%%%%%%%%%%%%%
% DMFT M1 full DOS %
%%%%%%%%%%%%%%%%%%%%
In the $M_1$ phase, the most prominent effect of the DMFT theory is
the appearance of a gap of $\sim 0.7$~eV at the chemical potential.
For both the rutile and $M_1$ phase the positions of the O-$2p$ and
the V-$3d-e_g$ states show good agreement with the experiments.  Our
calculated $Z$ factors for the $M_1$ phase are $\sim 0.9$ for the
$e_{g}^\pi$ state and $\sim 0.7$ for the $a_{1g}$ state confirming
that the $M_1$ phase is "less correlated" than the rutile phase, as
emphasized in Ref.~\onlinecite{Tomczak2}.
\begin{figure}[!bt]
\centering{
$\begin{array}{cc}
\includegraphics[height=0.3\linewidth]{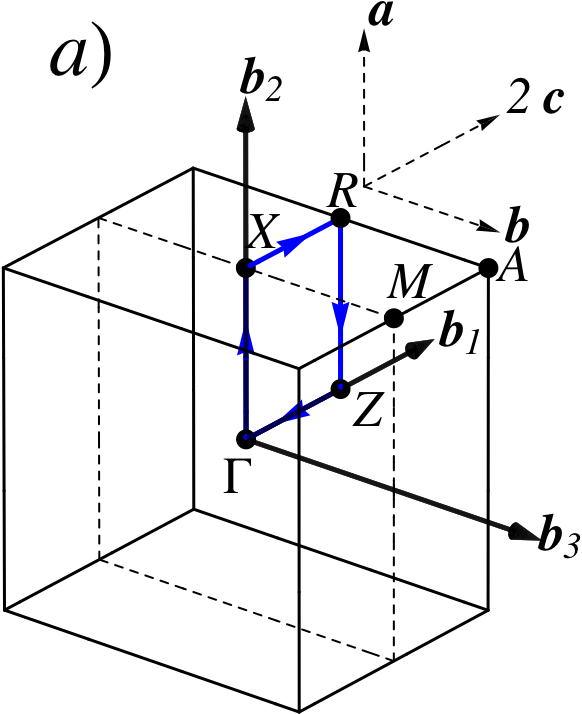}&
\includegraphics[height=0.3\linewidth]{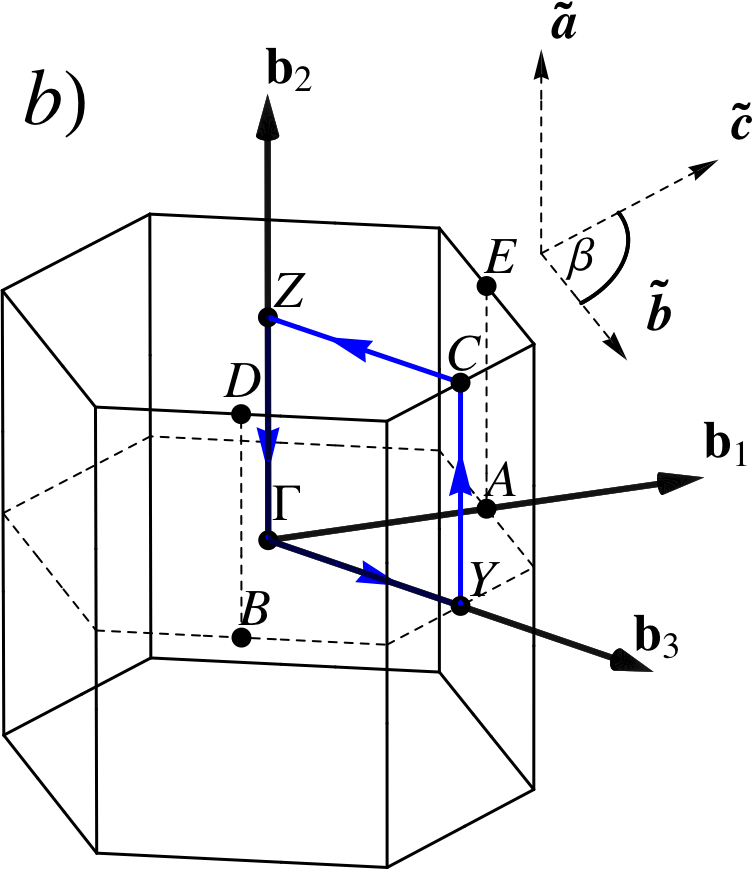}\\
\includegraphics[width=0.49\linewidth]{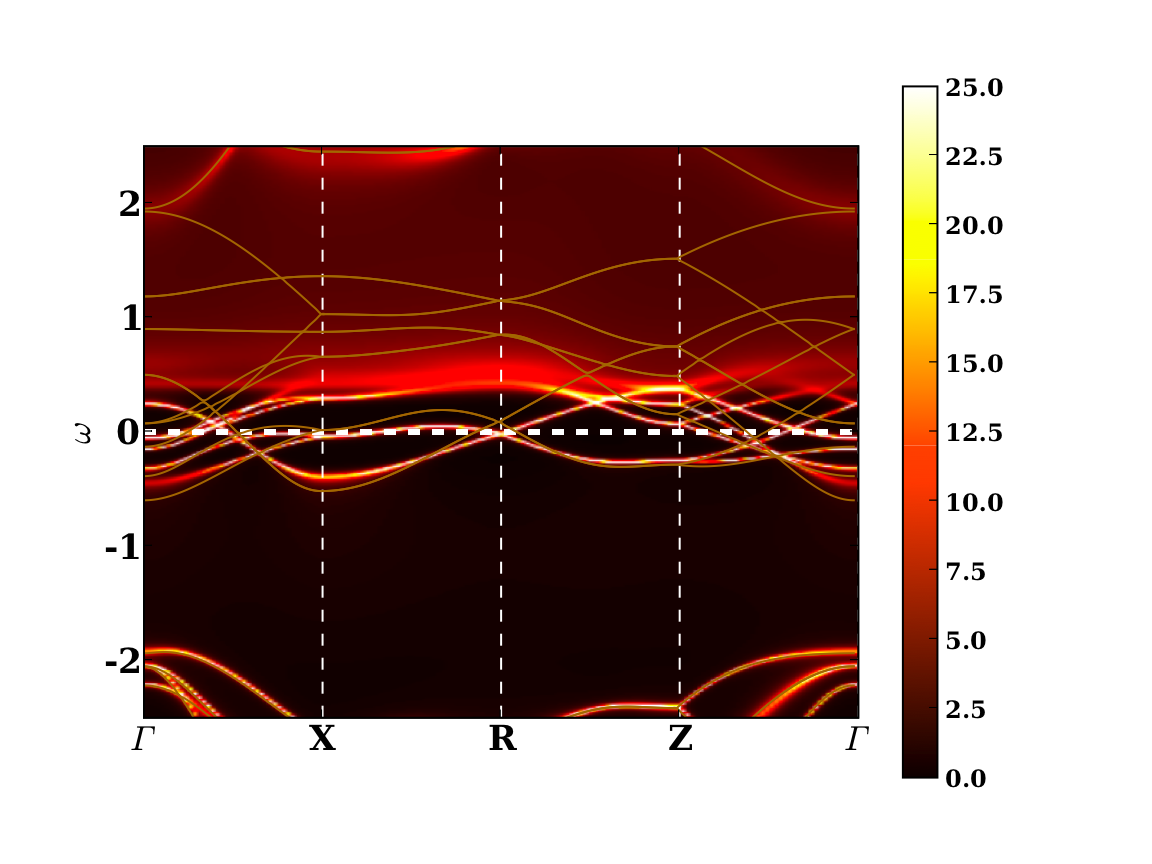}&
\includegraphics[width=0.49\linewidth]{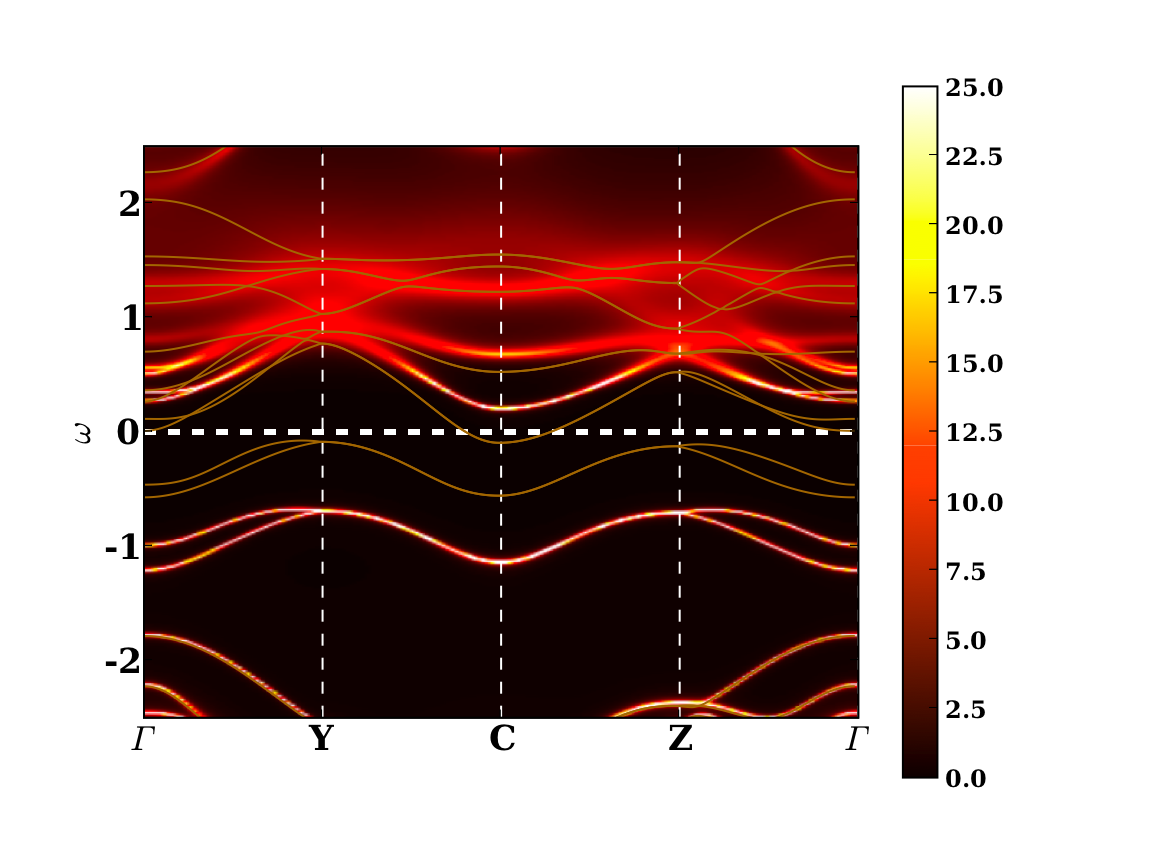}
\end{array}$\\
  }
\caption{
(Color online)
{\em Upper panel:} Brillouin zones and high symmetry points of
the rutile $(a)$ and the $M_1$ $(b)$ structures. 
{\em Dashed arrows:} the lattice vectors, 
{\em thick arrows:} the reciprocal lattice vectors, 
{\em dashed lines:} guide to the eye.
{\em Lower panel:} The momentum resolved spectral function 
$A(\vec{k},\omega)$ along the high
symmetry paths (shown in the Brillouin zones by blue lines) 
of the {\em R} (left) and $M_1$ (right)
phases.
For rutile the Brillouin zone corresponds to the doubled unit cell as 
indicated by the lattice vectors.
The full lines show the
LDA bands, and the color-coding shows the LDA+DMFT spectra.
$\omega$ is in units of eV.
}
\label{Akw}
\end{figure}

Fig.~\ref{Akw} shows the momentum resolved spectra in the
{\em R} and $M_1$-phase.  
Notice the downshift of the two bonding bands in the $M_1$ phase,
primarily of $x^2-y^2$ character (notice that there are four V
atoms per unit cell, and hence two bonding $x^2-y^2$ bands). The
states above the Fermi level move slightly up and shrink due to
many-body renormalization similarly to the $t_{2g}$ states in the {\em R} phase.
The bands around $1\,$eV also acquire substantial lifetime 
in both phases.
%
%The oxygen bands below $-1.5\,$eV and the $e_g$ bands above $2\,$eV are
The oxygen bands below $-1.5\,$eV and the $eg$ bands above $2\,$eV are
almost unchanged compared to LDA.

The results for the orbitally resolved $3d-t_{2g}$ spectral functions
are shown in Fig.~\ref{fig:resolved}.  In the rutile phase one can see
that the three $t_{2g}$ states are approximately equally occupied
predicting isotropic transport properties.  For the $a_{1g}$ state the
bonding anti-bonding structure can be also recognized similarly to the
$M_1$ phase with a splitting of $\sim0.4$~eV which is by a factor of 2
smaller then in the LDA results. The calculated positions of the
$t_{2g}$ states agree well with the XAS results.  In the $M_1$ phase,
the weight redistribution is very different: only a single state,
namely the bonding $a_{1g}$ orbital is occupied.  One can observe that
the bonding $a_{1g}$ state is shifted for $\sim0.8$~eV lower, which is
in good agreement with the previous theoretical results \cite{BPL+05}.
The spectral function at the upper edge of the gap has predominantly
the $e_g^\pi$ character, but there is also some weight of the
anti-bonding $a_{1g}$ character, which is in agreement with experiment
showing that the spectral density has not purely $e_{g}^\pi$ character
above the gap \cite{KHH+06}.
The first two peaks at $\sim0.6$~eV above the chemical potential are
attributed mainly to the $e_g^\pi$ states, and the third one is due to
the $a_{1g}$ state. This is consistent with the recent results of
polarization dependent O K XAS experiments from Koethe {\em at al.}
\cite{KHH+06} where the orbital character of the states can be deduced
by changing the polarization of the x-ray from parallel (O K XAS
$\parallel$) to the $c$-axis to perpendicular polarization (O K XAS
$\perp$).  The anti-bonding $a_{1g}$ state lies at 1.3~eV above the
Fermi level.  This peak did not appear in previous theoretical studies
of Ref~\onlinecite{BPL+05}, but agrees well with current XAS results,
which show that there is a prominent $a_{1g}$ $\sim1.0$~eV above the
$e_g^\pi$ peak \cite{KHH+06,AGF+91,SST+90}.  It is interesting to note
that the position of the anti-bonding $a_{1g}$ state is roughly the
same in both the DMFT results and the pure LDA results.
The calculated separation between the bonding and
anti-bonding peaks of the $a_{1g}$ state is $\sim 2.1$~eV, which
agrees reasonably with the experimentally found value
($2.5-2.8$~eV)\cite{KHH+06}. 
Finally, the peak around $3\,$eV of the XAS spectra may be assigned to
the contribution from the $e_g$ states, included in Fig.~\ref{fig:total},
but excluded in Fig.~\ref{fig:resolved}. This is consistent with the
experimental finding of Ref.~\cite{KHH+06} where negligible
change of the peak weight was observed across the MIT, but strong
sensitivity to the polarization was noticed.

\begin{figure}[th!]
\includegraphics[width=0.4\linewidth]{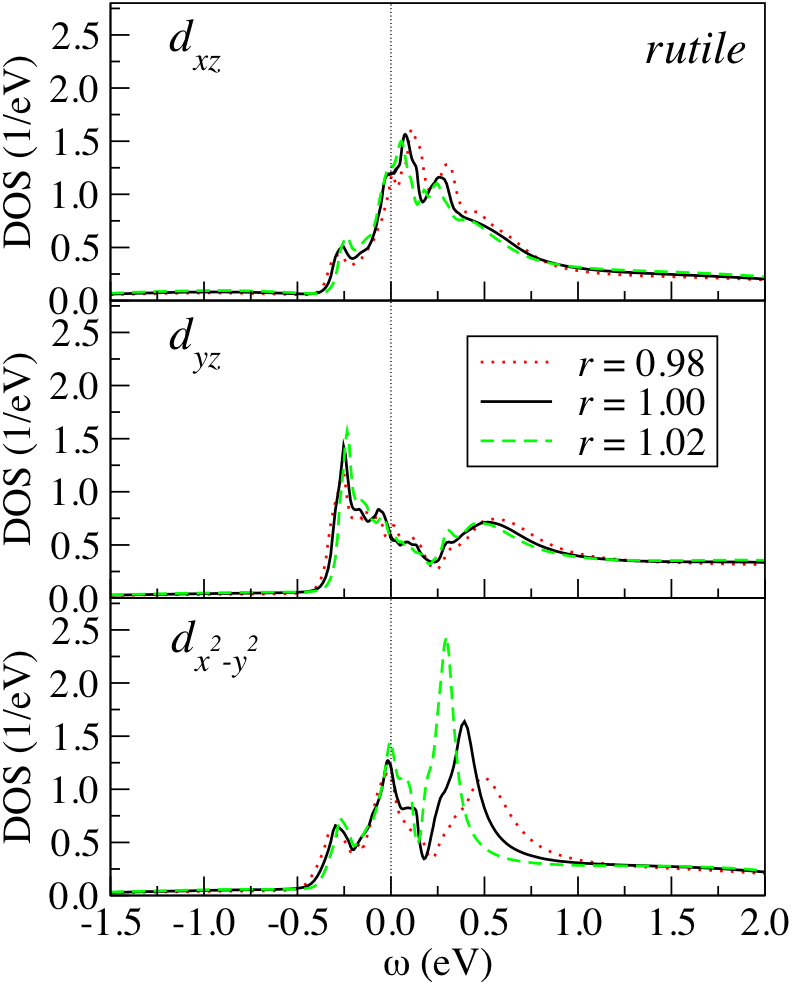}
\includegraphics[width=0.4\linewidth]{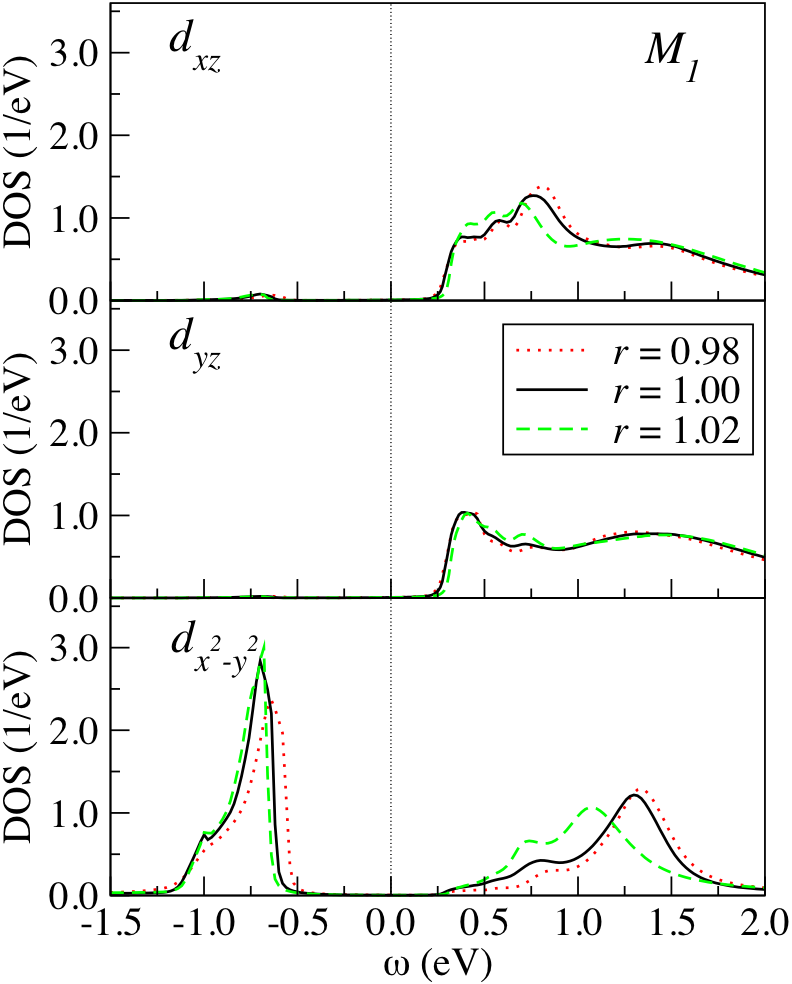}
\caption{
(Color online)
Orbitally resolved V$-3d-t_{2g}$ DOS 
of a V atom
in rutile (left) and $M_1$ (right) phases in case of $r=0.98$ (dotted, red line),
1.00 (full, black line) and 1.02 (dashed, green line).
}
\label{fig:dos_strain}
\end{figure}

%%%%%%%%%%%%%%%%%%%%%%%%%%%%
% DMFT rutile resolved DOS %
%%%%%%%%%%%%%%%%%%%%%%%%%%%%
To elucidate the effect of the strain along the $c$-axis
the spectral functions of the $t_{2g}$ orbitals for
different $r$ ratios ($r=0.98$, 1.00 and 1.02)
are shown in Fig.~\ref{fig:dos_strain} 
for both phases.
In the rutile phase the $e_{g}^\pi$ states are hardly affected by the
strain. In contrast, the bonding anti-bonding splitting of the 
$a_{1g}$ states  shows a strong sensitivity to strain.
The width of the upper $a_{1g}$ peak at $\sim0.5$~eV decreases 
with increasing $r$, which is due to the weaker hybridization 
between the V atoms, as indicated by the decreasing hopping integrals.
Surprisingly, the spectral weight at the chemical potential is practically 
unaffected by the changes of the lattice constant $c$.
% suggesting 
% that the transport properties are also hardly influenced
% but the anisotropy of transport should becomes more pronounced.
%
Similar calculation using cluster DMFT was carried out for the rutile
phase, allowing for the formation of split bonding-anti-bonding pairs
along the $c$-axis.  We did not find an appreciable sign for the
development of the bonding-antibonding splitting even for the $r=0.98$
case, confirming that the single site DMFT is quite accurate in the
rutile phase.

\begin{figure}[!bt]
\centering{
 \includegraphics[width=0.5\linewidth]{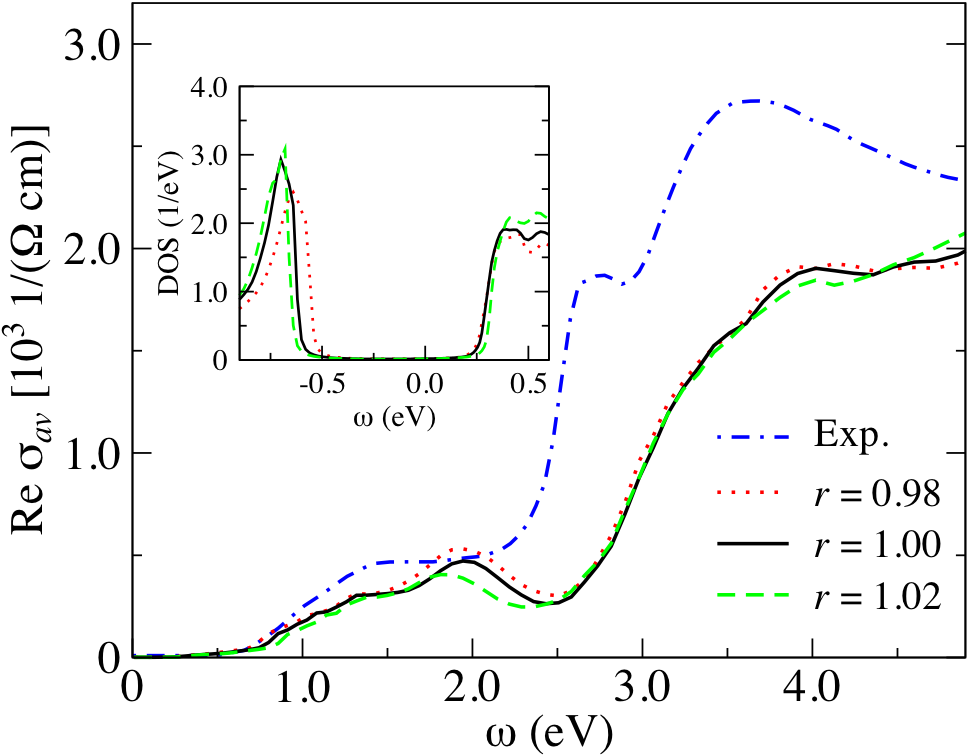}
}
\caption{
(Color online)
Real part of the averaged optical 
conductivity $\sigma_{av}=\frac{1}{3}(\sigma_{\parallel}+
2\sigma_{\perp})$ of the $M_1$ phase with different $r$ ratios.
Inset: total $t_{2g}$ density of states of the $M_1$ phase.
The experimental result is taken from Ref.~\cite{QBW+08}.
}
\label{fig:optics}
\end{figure}

%%%%%%%%%%%%%%%%%%%%%%%%%%%%
% DMFT M1 resolved DOS %
%%%%%%%%%%%%%%%%%%%%%%%%%%%%
%%% Stopped here!!!
%%%
Fig.~\ref{fig:dos_strain} clearly shows that the width of the bonding
$a_{1g}$ peak is increased with decreasing $r$ ratio, which can be
attributed to the increase in inter dimer hoppings.
For all ratios $r$, the gap in orbital space is indirect, i.e., the
valence band is of $a_{1g}$ character and the conduction band of
$e_{g}^\pi$ character. This is in agreement with the experimental
findings of Ref.~\onlinecite{KHH+06} demonstrating that the Peierls
physics is playing a secondary role in the gap opening in $M_1$ phase.
Due to the decreasing length of the $c$-axis, the $e_{g}^\pi$ states
are \textit{shifted to} slightly \textit{lower energy}, which together
with the broadening of the bonding $a_{1g}$ peak results in the
\textit{contraction} of the gap, despite the increase in the bonding
anti-bonding splitting of the $a_{1g}$ peaks.
The decrease of the gap size due to decreasing $c$-axis length is more
apparent in the inset of Fig.~\ref{fig:optics}, which shows the total
$t_{2g}$ DOS.
This result is supported by the experimental result of Muraoka and
Hiroi \cite{MH02} demonstrating that the decrease in lattice parameter
$c$ leads to decrease in the metal-insulator transition
temperature. This is a clear indication that smaller $c$-axis length
leads to a weakened stability of the insulator in the $M_1$ phase, and
consequently a smaller gap in $M_1$ phase.
This behavior is not expected for a Peierls type gap, which
increase as the lattice parameter decreases along the dimerized
chains.

The evolution of the gap is reflected also in the gap of the optical
conductivity \cite{millis04}.  In Fig.~\ref{fig:optics} the real part
of the average optical conductivity
$\sigma_{av}=\frac{1}{3}(\sigma_{\parallel}+ 2\sigma_{\perp})$ is
shown, where $\sigma_{\parallel(\perp)}$ is the optical conductivity
in the case where the polarization of incident light is parallel
(perpendicular) to the $c$-axis. The calculated optical gap and the
intensity of the first peak corresponding to the $t_{2g}-t_{2g}$
excitations compare well with the experimental results for
polycrystalline VO$_2$ films \cite{QBW+08}.  The shoulder around $\sim
2.5$~eV in the experimental optical conductivity, which is primarily
due to the inter-band transitions, is shifted slightly upwards ($\sim
0.5$~eV) in the theoretical result. This is an indication that the applied
downfolding method describes well the low-energy properties, but not
so well the higher energy interband excitations.

The fact that the gap in the density of states arises between orbitals
of different symmetry, indicates that the anisotropy of the transport
properties in this phase will be very sensitive to disorder and grain
boundaries which can drastically alter the orientation of these
orbitals changing the matrix elements for hopping and conductivity.

\begin{figure}[!bt]
\centering{
 \includegraphics[width=0.5\linewidth]{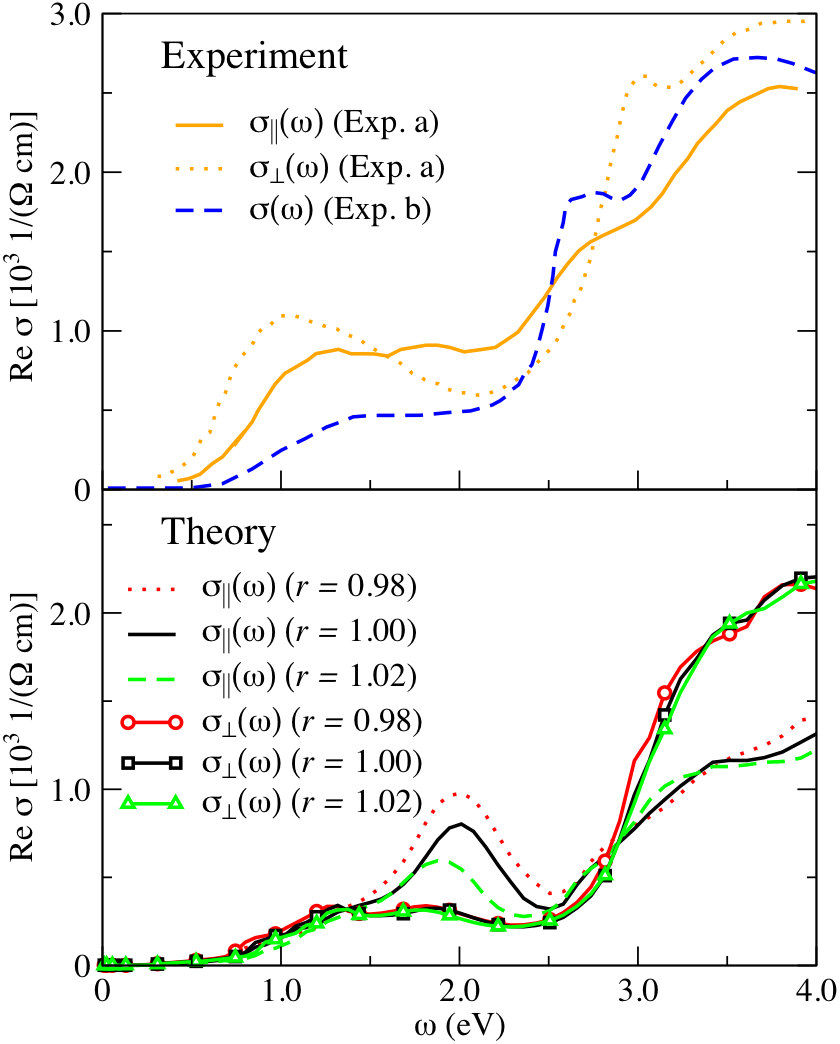}
  }
\caption{
(Color online)
Real part of the optical conductivity corresponding 
to light polarized parallel ($\sigma_{\parallel}$)
and perpendicular ($\sigma_{\perp})$ to the $c$-axis in  the $M_1$ phase with different $r$ ratios.
The experimental result is taken from Ref.~\cite{QBW+08} (Exp. a) and \cite{VBB68,TB09} (Exp. b).
}
\label{fig:optics_anis}
\end{figure}

Fig.~\ref{fig:optics_anis} shows the calculated and experimentally
measured optical conductivity for differently polarized light.  It can
be seen that the trends of the dependence of the optical conductivity
on the polarization are in a good agreement with experimental results,
although the values are slightly different.  When the polarization is
perpendicular to the $c$-axis the optical response is practically
unaffected by the strain. In case of parallel polarization the optical
conductivity is strongly modified by changing the lattice parameter
$\tilde{c}$, especially in the frequencies between $1.5-2.5$~eV. This
region can be attributed to the $d_{x^2-y^2}-d_{x^2-y^2}$ excitations
as can be concluded from the positions of the bonding and antibonding
peaks in Fig.~\ref{fig:dos_strain}.  This results strongly indicate
that the anisotropy of the transport properties is due to directed V-V
bonds along the $c$-axis.

\begin{figure}[!bt]
\centering{
 \includegraphics[width=0.5\linewidth]{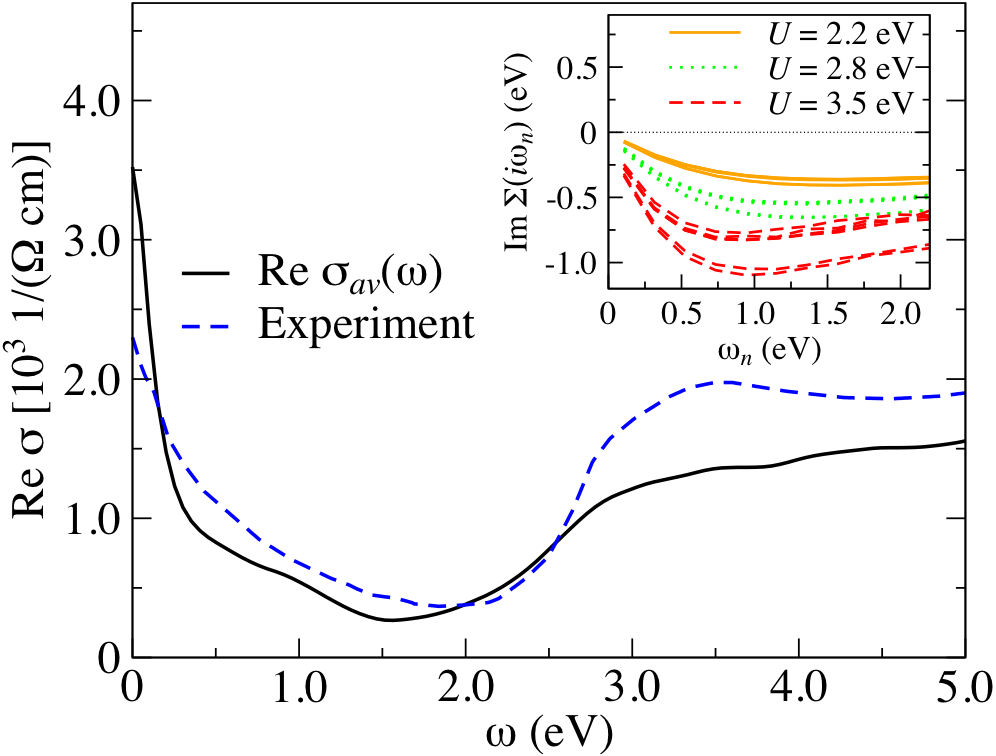}
}
\caption{
(Color online)
Real part of the averaged optical conductivity 
$\sigma_{av}=\frac{1}{3}(\sigma_{\parallel}+
2\sigma_{\perp})$ of the {\em R} phase.
The experimental result is taken from Ref.~\cite{QBW+08}.
{\em Inset:} Imaginary part of the self-energy corresponding to 
the $t_{2g}$ orbitals in the bonding anti-bonding basis at the imaginary 
Matsubara frequencies close to the real axis.
}
\label{fig:imsig_optics}
\end{figure}

In Fig.~\ref{fig:imsig_optics}, the orientation average optical
conductivity of the {\em R} phase and the imaginary part of the
self-energy at the Matsubara frequencies close to zero are shown.
While the area below the calculated and experimentally measured
\cite{QBW+08} optical conductivity (plasma frequency) agree fairly
well, the width of the two Drude peaks are different.  In
order to improve the agreement with experimental results \cite{QBW+08}
an imaginary part of 0.55~eV (scattering rate) was added to the
self-energy for the low frequency part of the optical conductivity, to
simulate the experimentally measured broadening of the Drude peak.
Inspecting the inset of Fig.~\ref{fig:imsig_optics}, one can see that
even by employing larger $U$ values, the scattering rate
($\mathrm{Im}\Sigma(\omega \rightarrow 0 )$) is not large enough to
reproduce the experimental results, and the calculated optical
conductivity will not show {\em bad metal} behaviour at this
temperature.  From this result one can draw the conclusion that the
experimentally measured large scattering rate is a consequence of an
inhomogeneity of the system as indicated in recent experiments
\cite{Qazilbash3,CES+09}.

\section{Phase diagram and limits of downfolding}

\begin{figure}[!bt]
\centering{
 \includegraphics[width=0.5\linewidth]{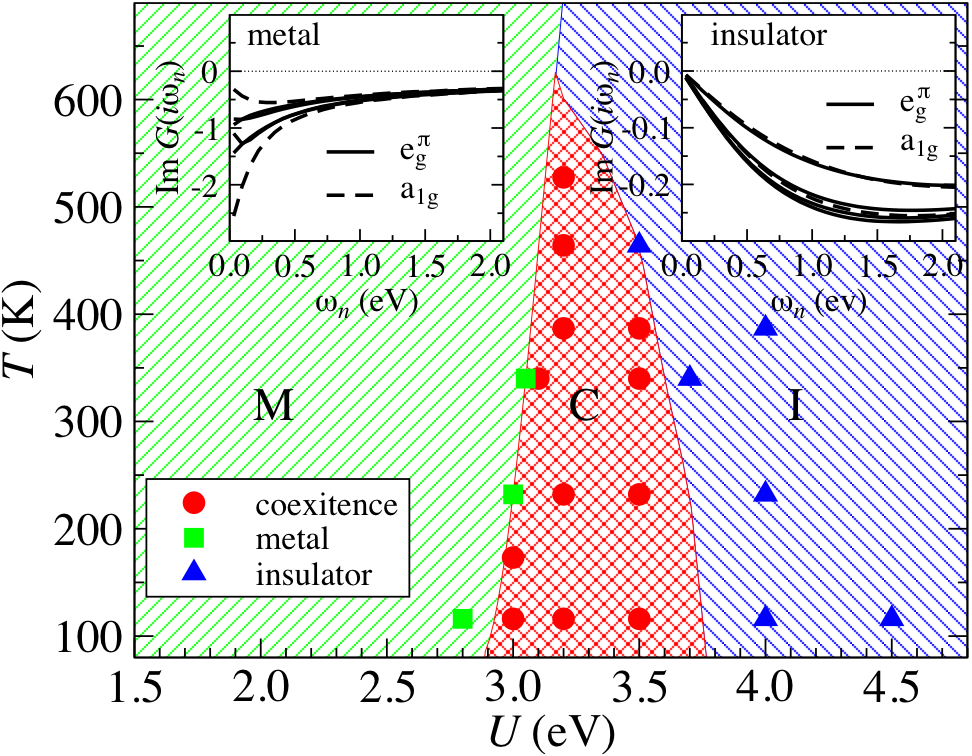}
  }
\caption{
(Color online)
Sketch of the phase diagram of the rutile phase as a function of 
the temperature and the applied Coulomb repulsion $U$ based on the 
calculated points (dots). 
In the green region (M) the metallic solution (shown in the left inset in 
terms of the Matsubara Green's function) 
is stable only 
while in the blue region only the insulator solution (right inset) can be found.
In the red coexistence region (C) both solutions are stable.
}
\label{fig:phase}
\end{figure}

In Fig.~\ref{fig:phase}, a sketch of the phase diagram of the rutile
structure based on calculated points within the cluster DMFT is
displayed as a function of the temperature and the Coulomb repulsion
parameter $U$.  Three different phases are distinguished: the metal
(M), the insulator (I) and the coexistent (C) regions.  To identify
them, the imaginary part of the local Matsubara Green's function,
$\mathrm{Im}\mathcal{G}(\omega_n)$, of the cluster was investigated.
We considered that the solution is insulating when
$\mathrm{Im}\mathcal{G}(\omega_n)$ converges to 0 at low imaginary
frequencies for all orbitals, as shown in an example in the right
inset of Fig.~\ref{fig:phase}.  The metallic solutions are the ones
where $\mathrm{Im}\mathcal{G}(\omega_n)$ tends to a finite value for
at least one of the orbitals (see the left inset in
Fig.~\ref{fig:phase}).  In the metallic region only the metallic
solution is stable. In the insulator region only the insulating
solution is stable, and in the coexistence region, both are stable.
In order to decide whether a solution is stable or not, the DMFT
calculations were started from an ansatz of a specified type (metal or
insulator) and if it remained of the same type after the
self-consistent solution is reached, one can regard it as a stable
mean-field solution.  In Fig.~\ref{fig:phase} it can be observed that
below $U\approx2.9$~eV only the metallic solutions are stable while
above $\sim3.8$~eV only the insulator is stable.
%\begin{color}{red}
The nature of the insulating state in rutile phase is very different
from the insulating state in $M_1$ phase. The lower Hubbard band
in the rutile insulator is an almost equal mixture of all three
$t_{2g}$ orbitals. However, the interaction strength needed to open
the true Mott gap without the help of the Peierls mechanism, is
considerably larger.
%\end{color}
%
In the coexistence region, one can expect a crossover between the two
phases governed by the free energy of the system \cite{PHK08}.  There
is a strong experimental indication that the rutile phase resides
in the vicinity (but on the metallic side) of this crossover. Pouget
and Launois showed that the metallic feature of the rutile phase is
very sensitive by substitutional alloying of VO$_2$ with Nb
(V$_{1-x}$Nb$_{x}$O$_2$) which increases the $c/a$ ratio and results
in the appearance of a gap keeping the rutile structure at $x=0.2$
\cite{JL76}.  Recently, Holman {\it et~al.} \cite{HMW+09} reported
insulator to metal transition in V$_{1-x}$Mo$_x$O$_2$ system at
$x\approx0.2$.  All those experiments suggest that VO$_2$ is in the
crossover region near the coexistence of two solutions in the cluster
DMFT phase diagram.
%
%According to our calculation, the insulating solution in the rutile
%phase, the interaction $U$ needs to be considerably stronger then the
%interaction strength $U$ which opens the gap in $M_1$ phase.
%

%\begin{color}{red}
On the basis of our calculation we conjecture that the rutile phase
might be able to support either metallic or insulating solution (with
a very small gap), and hence either of the two phases can be
stabilized depending on small external stimuli. On the other hand, the
$M_1$ phase at the same interaction strength supports only insulating
solution, and it is unlikely that small external perturbation can turn
it to metallic state.
%\end{color}

One possibility to improve the agreement between the experimental and
theoretical results is that one should use different parameters in the
downfolded model.  The second possibility is that the material is
strongly inhomogeneous which was not taken into account so far in any
theoretical calculations.  Finally, it is most likely that calculation
without invoking the downfolding approximation will result in a more
accurate description of this material. This would place the VO$_2$
closer to the Mott charge transfer insulator boundary in the
Zaanen-Sawatzky-Allen phase diagram.  Work in this direction is in
progress.

\section{Conclusions}

Our exploratory theoretical research set up the machinery for
describing the subtle interplay of Coulomb correlations, orbital
degeneracy and strain in determining the mechanism of the MIT in
VO$_2$. Our theory, coupled with existing strain experiments, clearly
shows that the Peierls distortion is only one element affecting the
MIT and the switching mechanism of this material.  The LDA+DMFT
calculations in the unstrained material are in good agreement with
experiments. We performed the first LDA+DMFT studies of the electronic
structure of VO$_2$ under strain. Besides the Peierls increase in $a_{1g}$
bonding-antibonding splitting, the lowering in energy of the $e^\pi_g$
orbital, and the rapid change in bandwidth of the $a_{1g}$ orbital due
to the varying overlaps, play an equally important role in controlling
the position of the MIT. These theoretical insights can be used for
understanding and improving material properties by means of chemical
substitutions.
For a more accurate description it is mandatory to take into account
the oxygen degrees of freedom, and calculations of the total energy
in the regions suggested by this exploratory work.

\section{Acknowledgments} 

This research has been supported by Grants No. 
DARPA W911NF-08-1-0203, 
NSF-DMR 0806937, and  
OTKA F68726.
We are grateful to Jan Tomczak for stimulating discussions.

\end{document}